\documentclass[12pt]{article}

\usepackage[margin=1in]{geometry}  % set the margins to 1in on all sides
\usepackage{graphicx}              % to include figures
\usepackage{amsmath}               % great math stuff
\usepackage{amsfonts}              % for blackboard bold, etc
\usepackage{amsthm}                % better theorem environments
\usepackage{url}        
\usepackage{booktabs}
\usepackage{subcaption}
\usepackage{pseudocode}
\newcommand{\bbeta}{\mbox{\boldmath$\beta$}}

\newcommand{\bmm}{\boldsymbol{m}}
\newcommand{\bx}{\boldsymbol{x}}

\newcommand{\by}{\boldsymbol{y}}

\newcommand{\bg}{\boldsymbol{g}}
\newcommand{\bu}{\boldsymbol{u}}
\newcommand{\balpha}{\boldsymbol{\alpha}}

\newcommand{\be}{\boldsymbol{e}}
\newcommand{\bzero}{\boldsymbol{0}}

\begin{document}

\title{Locally Epistatic Models for Genome-wide Prediction and Association by Importance Sampling}
\author{Deniz Akdemir\footnote{Corresponding author, e-mail: deniz.akdemir.work@gmail.com} \: \&  Jean-Luc Jannink \\ 
Plant Breeding and Genetics \\
Cornell University \\
Ithaca, NY 14853 USA}

\maketitle

\begin{abstract}
In statistical genetics an important task involves building predictive models for the genotype-phenotype relationships and thus attribute a proportion of the total phenotypic variance to the variation in genotypes. Numerous  models have been proposed to incorporate additive genetic effects into models for prediction or association. However, there is a scarcity of models that can adequately account for gene by gene or  other forms of genetical interactions. In addition, there is an increased interest in using marker annotations in genome-wide prediction and association. In this paper, we discuss an hybrid modeling methodology which combines the parametric mixed modeling approach and the non-parametric rule ensembles. This approach gives us a flexible class of models that can be used to capture additive, locally epistatic genetic effects, gene x background interactions and allows us to incorporate one or more annotations into the genomic selection or association models. We use benchmark data sets covering a range of organisms and traits in addition to simulated data sets to illustrate the strengths of this approach. The improvement of model accuracies and association results suggest that a part of the "missing heritability" in complex traits can be captured by modeling local epistasis. 
\end{abstract}

\section{Introduction}

There has been a great interest in explaining the total variation observed in a trait over individuals of a population. In general models are constructed such that the total variance is partitioned into the sum of a genetic component, an environmental component, and a component for the residual unexplained variance. Further refinements, might include addition of a component for genotype by environment interactions.

For complex quantitative traits usual assumption about the trait architecture is the classical infinitesimal model, introduced by \cite{fisher1918correlation}, where the genetic values (GVs) of individuals are assumed to be generated by an infinite number of unlinked and nonepistatic genes, each with an independent infinitesimal effect. This model is also called the polygenic model, it was developed into a sophisticated theory in the 1950’s (\cite{henderson1953estimation, cockerham1954extension, anderson1954model, kempthorne1954correlation}), and it has long been central to practical breeding where it forms the genetic basis for the animal model. This line of thought followed mainly from Mendel's observations that inheritance was discrete and discontinuous. In this context, the variances of phenotypes are described in terms of additive, dominance and epistatic components. The evidence from empirical studies of genetic variance components shows that additive variance  usually accounts for most of the total genetic variance (\cite{hill2008data, Huang041434}).

At the beginning of the 20th century, Thomas Hunt Morgan's showed that the genes responsible for the appearance of a specific phenotype were located on chromosomes and genes on the same chromosome do not always assort independently. This suggested that the strength of linkage between genes depended on the distance between them on the chromosome. The nearer two genes lie on a chromosome, the greater the chance of being inherited together. Likewise, the farther away they are from each other, the more chance of being separated by the process of crossover. This view was eventually captured by the double helix model for DNA of James Watson and Francis Crick in the 1950's (\cite{watson1953molecular}).

The DNA marker data available today for many model and non-model species provide a way to capture the polygene, and the infinitesimal model has been useful as a tool for detecting main effect loci by associating the common phenotypes with the common genotypes in the sample data. In addition, the genomewide predictive models mainly used genomic selection for breeding animals or plants showed that the results from models that assume additive infinitesimal effects are quite accurate and informative.   The infinitesimal model has very powerful simplifying statistical properties and avoids the need to specify individual gene effects. However, despite large sample sizes and increased number of markers and numerous statistical modeling approaches in the additive genetic framework, the 'missing heritability' problem still persists. Some of this unaccounted variance is believed to arise from a large number of loci with small individual contributions, or be due to epistasis and quite likely involves both effects.

Association studies of interaction among loci are complicated by the vast number of possibilities one has to consider.  In most association studies and models the focus is on estimating the effects of each marker and lower level interactions (\cite{cantor2010prioritizing}). For a marker data set of $m$ markers, a genome-wide only two way interactions analysis involving upto two loci will involve evaluating a number possibilities of the order of $m^2$ and $m$ can easily exceed millions. The methods used in identification and modeling epistasis usually lack statistical power, and they are computationally exhaustive or perhaps even unfeasible.

Most commonly used approach to reduce the problem dimension is to take a two step approach where the tested interactions are restricted to the markers which have significant additive effects. This approach, although may be viable, ignores the fact that epistatic effects does not need to be visible to methods that are built to catch only additive effects. In addition, the methods used for the second step are usually too restrictive in their forms. For example, a very common approach is to use a multiplicative term for two markers coded as 0, 1 or 2 (or any other coding based on allele frequencies).  This approach is not satisfactory (See table \ref{tab:interactions}). Adding additional terms other additive effects and their interactions, for example, including dominance terms can not fix this problem, mainly because the increase in the dimension, complexity and the confounding between these terms cause the estimates have high sampling error and there is little power to distinguish between the components. 

The model of epistasis based on variance components like dominance, additive x additive, additive x dominance, dominance x dominance, etc,... can not represent all kinds of genetic epistasis. For example, we might expect some genes behave differently in different genetic backgrounds, or genes acting in a hierachical network. There is no room for these kind of epistatic behaviour in classical quantitative genetics. However, there are numerous research about biological pathways and gene networks that indicate some genetic variance in populations is due to such interactions (\cite{schadt2005integrative, phillips2008epistasis, mackay2014epistasis, wei2014detecting}).

\begin{table}[htb]
\centering
    \begin{tabular}{rrrrrrrrrr}
    \toprule
    \textbf{Genotype-Phenotype} & \textbf{} &       &       &       &       & \textbf{Allele coding and m1*m2} & \textbf{} &       &  \\
    \midrule
    \multicolumn{1}{c}{\textbf{m1\textbackslash{}m2}} & \multicolumn{1}{c}{\textbf{BB}} & \multicolumn{1}{c}{\textbf{Bb}} & \multicolumn{1}{c}{\textbf{bb}} & \multicolumn{1}{c}{\textbf{}} & \multicolumn{1}{c}{} & \multicolumn{1}{c}{\textbf{m1\textbackslash{}m2}} & \multicolumn{1}{c}{\textbf{0}} & \multicolumn{1}{c}{\textbf{1}} & \multicolumn{1}{c}{\textbf{2}} \\
    \multicolumn{1}{c}{\textbf{AA}} & \multicolumn{1}{c}{+} & \multicolumn{1}{c}{+} & \multicolumn{1}{c}{+} & \multicolumn{1}{c}{} & \multicolumn{1}{c}{} & \multicolumn{1}{c}{\textbf{0}} & \multicolumn{1}{c}{0} & \multicolumn{1}{c}{0} & \multicolumn{1}{c}{0} \\
    \multicolumn{1}{c}{\textbf{Aa}} & \multicolumn{1}{c}{-} & \multicolumn{1}{c}{+} & \multicolumn{1}{c}{+} & \multicolumn{1}{c}{} & \multicolumn{1}{c}{} & \multicolumn{1}{c}{\textbf{1}} & \multicolumn{1}{c}{0} & \multicolumn{1}{c}{1} & \multicolumn{1}{c}{2} \\
    \multicolumn{1}{c}{\textbf{aa}} & \multicolumn{1}{c}{-} & \multicolumn{1}{c}{+} & \multicolumn{1}{c}{+} & \multicolumn{1}{c}{} & \multicolumn{1}{c}{} & \multicolumn{1}{c}{\textbf{2}} & \multicolumn{1}{c}{0} & \multicolumn{1}{c}{2} & \multicolumn{1}{c}{4} \\
    \bottomrule
    \end{tabular}% 
  \caption{A scenario which shows an interaction pattern between two markers generated by a simple rule  ''$I(m1<2)*I(m2>1)\rightarrow$ - (, else +)''. The standard multiplicative formulation ($m1*m2$) cannot adequately represent this interaction and other terms would be needed in the model (additive, additive*additive, additive*dominance, dominance*dominance, see factorial model in \cite{anderson1954model, kempthorne1954correlation}).}
	\label{tab:interactions}
\end{table}

Some methods for capturing genome-wide epistasis include the RKHS regression approach and related support vector machines regression and the partitioning based random forest. These models can be used to predict the genetic values. However, these methods do not provide satisfactory information about genetic architecture of traits.   In addition, from the point of the breeder, it is not possible to know how much of this accuracy gain can be passed onto new generations.  These models do not distinguish between local and genome-wide interactions. 

An alternative approach for reducing the dimension of the problem when studying epistasis is considering only local epistasis (\cite{akdemir2015locally}), i.e., only epistatic interactions between closely located alleles  It is reasonable to assume that only epistatic effects that arise from alleles in gametic disequilibrium among closely located loci can contribute to long term response since there is a constant competition between epistatic selection and recombination.  In the presence of epistasis, selection, by increasing the frequency of favorable genotypes, establishes correlations between alleles at different loci and functionally related genes tend to cluster (16, 17), suggesting selection on gene order.  Furthermore, chromosomes have regions of infrequent recombination, interspersed with recombination hot-spots (18).

A mathematical argument for focusing on short segments of the genome as distinct structures comes from the ''building blocks'' hypothesis in the evolutionary theory. For instance, the schema theorem of Holland (\cite{holland1975adaptation}) predicts that a complex system that uses evolutionary mechanisms such as fitness, recombination and mutation tend to generate short and well fit and specialized structures whose number will increase exponentially in successive generations. For example, when the alleles associated with an important fitness trait are scattered all around the genome the favorable effects can be lost by independent segregation. Therefore, inversions that group these alleles physically together would be selected.  This is the basis for the observation that short (defining length), low order schema of
 above average population fitness will be favored. The effects that are selected over a long time scale will be those that can be broken down into useful parts. Said in another way, a beneficial epistatic effect with a short defining length is more fit than an epistatic effect with a longer defining length with the same effect.

In this article, we propose a hybrid (machine learning + mixed models) approach gives us a flexible class of models that can be used to capture additive, locally epistatic genetic effects, gene x background interactions and allows us to incorporate one or more annotations into the genomic prediction and association models.  A main aim of this article is to measure and incorporate additive and local epistatic genetic contributions since we believe that the local epistatic effects are relevant to the breeder. Another important point of the article makes is that the locally epistatic framework simplifies the study of interactions. 

The rest of the article is organized as follows. The next section provides a review of the most commonly used prediction and association models. In section 3, we introduce the locally epistatic models we use in this article. Prediction and association with these models is explained here. Section 4, is the examples section, where real and simulated data-sets are used to illustrate the proposed methodology and provide comparisons. We conclude the paper with a summary of findings, comments and future directions for research.

\section{Methods}

\begin{figure}[htb]
\centering
 
    \includegraphics[width=1\textwidth, angle=0]{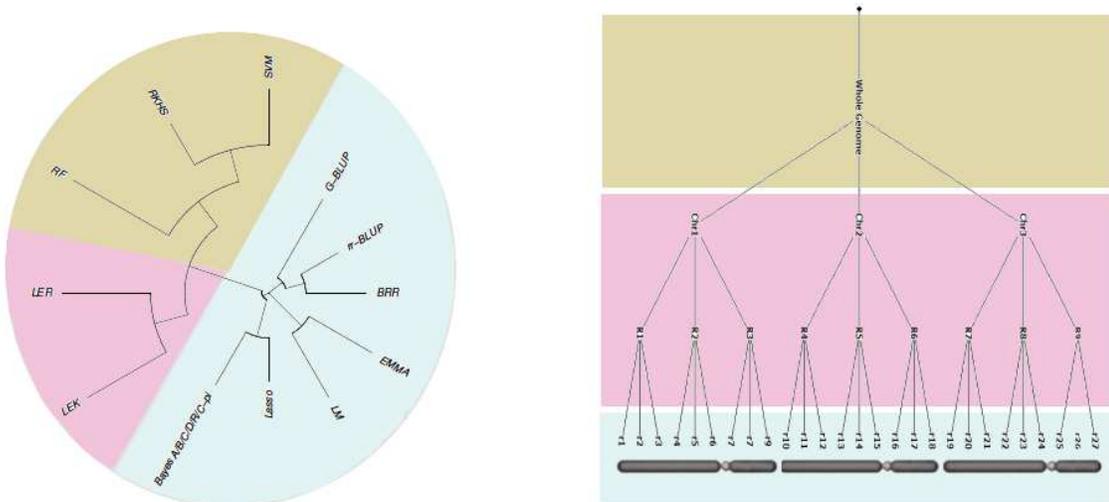} 
  \label{fig:fig1}
  \caption{Many of the models used in genomic prediction and association are additive: This include Ridge regression-Best Linear Unbiased Prediction (rr-BLUP) (\cite{whittaker2000marker, meuwissen2001prediction}), Lasso \cite{tibshirani1996regression}, Bayesian-Lasso (\cite{park2008bayesian}), Bayesian ridge regression, Bayesian alphabet (\cite{gianola2009additive, sorensen2007likelihood}), G-BLUP , EMMA (\cite{zhou2012genome}).  Several scientists have also developed methods to use genome-wide epistatic effects: RKHS (\cite{gianola2008reproducing, de2009reproducing}), RF (\cite{breiman2001random}), SVM. }
	\label{fig:fig2}
\end{figure}

There are numerous statistical models used in genomic prediction and association (see Figure \ref{fig:fig2}).  An evaluation of these methods for prediction of quantitative traits can be found in Heslot (2011).  In the rest of this section, we will briefly describe some of these model since they are important for developing the methodology in this paper.

Mixed models (MM) methodology has a special place in quantitative genetics because it provides a formal way of partitioning the variability observed in traits into heritable and environmental components, it is also useful in controlling for population structure and relatedness for genome-wide association studies (GWAS). In a mixed model, a genetic information in the form of a pedigree or markers can be used in the form of an additive genetic similarity matrix that describes the similarity based on additive genetic effects (G-BLUP). For the $n\times 1$ response vector $\by,$ G-BLUP model can be expressed as 
\begin{equation}\label{eq:gblup} \by=X\beta+Z\bg+\be \end{equation} where $X$ is the $n\times p$ design matrix for the fixed effects, $\beta$ is a $p\times 1$ vector of fixed effect coefficients, $Z$ is the $n\times q$ design matrix for the random effects; the vector random effects $(\bg',\be')'$ is assumed to follow a multivariate normal (MVN) distribution with mean $\bzero$ and covariance \[\left( \begin{array}{cc}
\sigma^{2}_{g} G  & \bzero  \\
\bzero & \sigma^{2}_{e} I_{n} \end{array} \right)\] where $G$ is  the $q\times q$ additive genetic similarity matrix. Given $M,$ the marker allele frequency centered incidence matrix, the matrix $G$ can be calculated as $G= {MM'/k}$ where $k$ is twice the sum of heterozygosities  of the markers (VanRaden, 2008).

It is known that the model (\ref{eq:gblup}) is equivalent to a MM in which the additive marker effects are estimated via the following model (rr-BLUP):
\begin{equation}\label{eq:rrblup} \by=X\bbeta+ZM\bu+\be \end{equation}  where $X$ is the $n\times p$ design matrix for the fixed effects, $\beta$ is a $p\times 1$ vector of fixed effect coefficients, $Z$ is the $n\times q$ design matrix for the random effects $M$ is $q\times m$ marker allele frequency centered incidence matrix; $(\bu',\be')'$ follows a MVN distribution with mean $\bzero$ and covariance \[ \left( \begin{array}{cc}
\sigma^2_u I_m  & \bzero  \\
\bzero & \sigma^2_e I_n \end{array} \right).\]   

RKHS model regression extends additive G-BLUP model by allowing a variety of similarity matrices, not necessarily additive in the input variables, calculated using a variety of kernel functions. Some common choices include the polynomial, and Gaussian kernel matrices (\cite{scholkopflearning}). It is possible to construct kernel matrices based on only dominance terms and other alternative codings of the markers.  In addition, the information in two or more such genome-wide kernels can be combined in a variance component model.  One problem with such approach is that these models cannot be used to learn the genetic architecture of the trait since kernel matrices measuring genome-wide similarity will in general be confounded, i.e., the estimates of variance components do not represent the contribution of individual terms (\cite{Huang041434}). 

The epistatic effects involving unlinked loci have high probability of being lost due to recombination, they will not contribute to subsequent response. Therefore, joint consideration of  linkage and epistasis is a necessary step for the models incorporating the interactions of more than one locus. In a recent article (\cite{akdemir2015locally}), we have proposed a modeling approach that uses RKHS based approach to extract the locally epistatic effects, which we refer to as the locally epistatic kernels (LEK) model. Briefly, the fitting procedure for LEK can described by the following steps: 
\begin{itemize}
\item Extract locally epistatic effects: \begin{equation}\label{eq:spmmmk3} \by=X_j\bbeta+Z\bg_j+\be_j, \end{equation} where  $\bg_j\sim N_{q_k}(\bzero, \sigma^2_{g_j}K_j)$ for $j=1,2,\ldots,k,$ $\be_j\sim N_{n}(0,\sigma^2_{e_j} I)$ and   $\bg_j,$ $\be_j$ are independent.
\item Estimate the coefficients of the following additive model:
\begin{equation}f(\bx,\bmm; \bbeta, \alpha)=\beta_0+\sum_{j=1}^{k}\alpha_{j}\hat{g}_{j}+\sum_{j=k+1}^{k+p}\beta_j x_j.\label{eq:additivemodel}\end{equation}
\end{itemize}

It was shown in  \cite{akdemir2015locally} that LEK models could be used to improve prediction accuracies and provide useful information about the genetic architecture. 

\subsection{Locally epistatic models via rules (LER)}

Ensemble learning provides solutions to complex statistical prediction problems by simultaneously using a number of models  (\cite{ho1990combination},  \cite{hansen1990neural},  \cite{kleinberg1990stochastic}, \cite{breiman1996bagging}),  (\cite{freund1996experiments}).  Random Forests (RF) (\cite{breiman2001random}) is a popular ensemble learning approach which also found its way to genomic prediction (\cite{heslot2012genomic}). Random forests is an ensemble of regression or decision trees which are obtained by re-sampling the data and the input variables. The nodes of a tree gives a partitioning of the input variables and the indicator function of each of these partitionings is called a rule. Rules can be used as input variables in regression or classification that gives rise to rule ensembles (\cite{friedman2008predictive, seni2010ensemble, akdemir2014ensemble}). 

%%%%%

A tree with $K$ terminal nodes define a $K$ partition of the input space where the membership to a specific node, say node $k,$ can be determined by applying the conjunctive rule $r_k(\bx)=\prod_{l=1}^{p}I(x_l\in s_{lk}),$ where $I(.)$ is the indicator function, $\bx=(x_1,x_2,\ldots, x_p)$ are the input variables. The regions $s_{lk}$ are intervals for a continuous variable and a subset of the possible values for a categorical variable. The complexity of trees or rules  (the degree of interactions between the input variables) in the ensemble increases with the increase in number of nodes from the root to the final node (depth). An ensemble of rules can be extracted from an ensemble of trees which can be generated using any of the standard Bagging, Random Forest, AdaBoost, and Gradient Boosting algorithms which are special cases of the importance sampling learning ensembles (ISLE) model generation procedure (\cite{ friedman2008predictive,seni2010ensemble}).

Suppose we have $n$ observations of the response variable written in a vector $\by=(y_1,$ $y_2,$ $\ldots,$ $y_n)'.$ Also let $X$ the $n\times p$ matrix of corresponding input variables. We would like to find a function of the p-dimensional input variables say $\bx$ that estimates the response variable $y.$  The pseudo code to produce $M$ base learners $\{f(\bx, \widehat{\theta}_j)\}_{j=1}^{M}$ under ISLE framework is given Algorithm \ref{ISLE}. $L(.,.)$ is a loss function; $S_j(\eta)$ is a subset of the indices $\{1,2,\ldots, n\}$ chosen by a sampling scheme $\eta,$ and $0\leq \nu \leq 1$ is a memory parameter.  We have a $p$ vector of input variables $\bx$ and a model family $F=\{f(\bx, \theta): \theta \in \Theta\}$ indexed by the parameter $\theta.$ The final ensemble models considered by  the ISLE framework  have an additive form: $F(\bx)=w_0+\sum_{j=1}^{M}w_{j} f(\bx, \theta_j)$
where $\{f(\bx, \theta_j)\}_{j=1}^{M}$ are base learners selected from the model family $F.$ Therefore, ISLE approach produces a generalized additive model (gam) (\cite{hastie1986generalized}).

ISLE uses a two-step approach to produce $F(\bx)$. The first step involves sampling the space of possible models to obtain $\{\widehat{\theta}_j\}_{j=1}^{M}.$ The space of models is usually sampled by sampling the instances and input variables and finding the best model in a predefined class of models $F$ for this subset of the data. The second step proceeds with combining the base learners by choosing weights $\{w_j\}_{j=0}^{M}.$  

\begin{pseudocode}{ISLE}{M, \nu, \eta}
\label{ISLE}

$$F_0(\bx)=0.$$ \\
\FOR $j=1$ \TO $M$ \DO
\BEGIN
$$(\widehat{c}_j, \widehat{\theta}_j)= \underset{(c,\theta)}{\operatorname{argmin}}\sum_{i \in S_j(\eta)} L(y_i, F_{j-1}(\bx_i)+cf(\bx_i, \theta))$$ \\
$$T_j(\bx)=f(\bx, \widehat{\theta}_j)$$ \\
$$F_j(\bx)=F_{j-1}(\bx)+\nu\widehat{c}_j T_j(\bx)$$ \\
\END \\
\RETURN{$$\{T_j(\bx)\}_{j=1}^M$ and $F_M(\bx).$$}
\end{pseudocode}
  
The \bfseries rulefit \normalfont algorithm of Friedman \& Popescu \cite{friedman2008predictive} uses an ensemble of rules (using trees as base learners) and a glmnet based post-processing step to calculate the  weights of the rules in an additive model. A few other post-processing approaches like partial least squares regression, multivariate kernel smoothing and weighting as well as use of rules in semi-supervised and unsupervised learning were described in \cite{akdemir2014ensemble}.

Locally epistatic rule based model fitting starts with definition of regions, suppose we defined $k$ such regions. This is followed by extraction of local rules from each genomic region $j=1,2,\ldots, k.$ using the ISLE algorithm. The rules are extracted from trees that predict the estimated genetic value from markers in the region.  Since the rules are independently generated for each region, this step can be computationally accomplished in parallel without loading the whole genetic data to computer RAM. The values of the rules from all regions are calculated for the $n$ training individuals, they are standardized with respect to their sample standard deviation and combined in a matrix  $n\times r$ matrix $R.$   

The second step in locally epistatic rule based model fitting is the post-processing step where we obtain a final prediction model using the extracted rules as input variables. In this article, we use the rr-BLUP model for post-processing the rules: \begin{equation}\label{eq:spmm2} \by=X\bbeta+Z R\balpha+\be, \end{equation} 
 where $n\times q$  is design matrix for the random effects, $R$ is  $q\times r$ design matrix for the centered and scaled rules, and $(\balpha',\be')'$ follows a MVN distribution with mean $\bzero$ and covariance \[ \left( \begin{array}{cc}
\sigma_{\balpha}^2 I_{r}  & \bzero  \\
\bzero & \sigma^2_e I_n \end{array} \right).\]    

Note that each rule is a function of the markers. Using estimated coefficients,  $\widehat{\balpha},$   we calculate the estimated genotypic value for an individual with markers  $\bmm$ as $\widehat{R(\bmm)}\widehat{\balpha}$  where $R(\bmm)=(R_1(\bmm),R_2(\bmm),\ldots,R_r(\bmm)).$  

In addition to having good prediction performance, a good model should also provide a description of the relationship between the input variables and the response. The rules and the estimated coefficients of the LER model can be used extract several importance and interaction measures. Let  $I(m_\ell\in R_j)$ denote the indicator function for the inclusion of marker $M_\ell$ in rule $R_j.$

\begin{itemize}

\item Since $R(\bmm)$ has standardized columns, $|\widehat{\alpha}|$ can be used as importance scores for the rules in the model.
\item A measure of importance for a marker $\ell$ is obtained by $I_j=\sum_{j=1}^{r}|\widehat{\alpha_j}| I(m_\ell\in R_j).$
\item A measure of interaction strength between two markers $\ell$ and $\ell'$ is obtained by: $I_{\ell \ell'}=\sum_{j=1}^{r}|\widehat{\alpha_j}| I(m_\ell\in R_j)I(m_{\ell'}\in R_j).$
\item A measure of interaction strength between markers $\ell_1, \ell_2,\ldots, \ell_l $ is given by \\ $I_{\ell_1 \ell_2\ldots,\ell_l}=\sum_{j=1}^{r}|\widehat{\alpha_j}| \prod_{k=\ell_1}^{l}I(m_{\ell_k}\in R_j).$
\item Importance of a region: Sum of the rule or marker importances within a region.
\end{itemize}

If environmental covariates are  observed along the  trait values then it is possible to include these variables with the markers in each region while extracting rules. This will allow environment main effects + gene by environment interaction terms enter the model. Variables measuring  background genetic variability related to the structure of the population can be incorporated in the model the same way. We note, however, that the importance and interaction measures for these variables will be inflated compared to that of the markers by a factor of the number of regions in the model. In the examples below, we have used the first three principal components of the marker matrix along with the markers to account for the genome-wide structural effects + gene interactions. It would be easy to extend this approach to include a hierarchy of interactions from genome-wide to SNPs. 

The depth of a rule is a parameter of the LER models since it controls the degree of interactions. A term involving the interaction of a set of variables can only enter the model if there is a rule that splits the input space based on those variables. One way to control the amount of interactions is to grow the trees to a certain depth. We can call this parameter the ''maxdepth'' parameter. In this article, we have allowed different rules enter the model by setting the ''maxdepth'' of each tree independently to a random variable generated from truncated Poisson distribution which turned the parameter into a continuous one which controls the "mean depth" of rules. This allows a diverse set of rules with different depths. In addition, trees and the associated rules can be pruned during extraction with heuristics like complexity cost pruning, or reduce error pruning. 

After extracting rules from a region a variable selection procedure can be applied to pick the most relevant rules from that region. A regression of the response variable on the set of rules from a region using the elastic-net loss function allows us to control the  number of rules selected as relevant from that region. In particular, elastic-net algorithm uses a loss function that is a weighted version of lasso and ridge-regression penalties. If all weight is put on the ridge-regression penalty no selection will be applied on the input variables. On the other extreme,  if all weight is put on the lasso penalty this will give maximal sparsity. We have treated this parameter as an hyper-parameter.  The remaining parameters of the elastic-net regression were selected using cross validation.

While fitting the model in (\ref{eq:additivemodel}), we need to decide on the values of a number of hyperparameters. Apart from the model set-up that involve the definition of genomic regions, and inclusion  or exclusion of some environmental or structural covariates; these parameters are ''mean depth'' parameter, number of rules extracted from each region (''nrules''), the proportion of markers (''proprow'')  and examples to use (''propcol''), parameters related to tree pruning and the parameters related to the elastic-net used in the filtering step. 

The hyper-parameters in LER models may be selected by comparing the cross-validated accuracies within the training data set for several reasonable choices.  Aggressive use of cross-validation can be supported by the theorem in \cite{vaart2006oracle}. However, the hyper-parameter choice for the LER models should also reflect the available resources and the needs. For instance, the number of regions that we can define depends on the number of markers and the the resolution the  data-set allows, and a more detailed analysis might only be suitable when the number of markers and the number of genotypes in the training data set are large. The hyper-parameters of the shrinkage estimators in the filtering step allows us to control the sparsity of the model. These parameters can be optimized for accuracy using cross-validation, but their value can also be influenced by the amount of sparsity desired in the model.  LER methodology provide the user with a range of models with different levels of detail, sparsity,  interactions.

\section{Examples}

%------------------------------------------------

We used four real life data sets to compare the prediction accuracies of the LER models to the standard linear G-BLUP model. For a few selected instances we also provided the association results. 

The data sets used in this article are summarized in Table \ref{tab:tab1}. The maize data set is taken from \url{panzea.org}, and was used in several articles (\cite{peiffer2014genetic, glaubitz2014tassel}). The rice dataset can be downloaded from \url{www.ricediversity.org} and was used in \cite{spindel2015genomic, begum2015genome}. Mouse dataset has been published in \cite{valdar2006genetic}, we have accessed this data from the synbreeddata package (\cite{synbreed}) available in R (\cite{Rmanula}).  We have downloaded the wheat dataset from \url{www.triticaletoolbox.org}. The curated versions of these datasets are also available from the corresponding author on request.

\begin{figure}[htb]
\centering
  \begin{tabular}{@{}cc@{}}
    \includegraphics[width=.4\textwidth, angle=270]{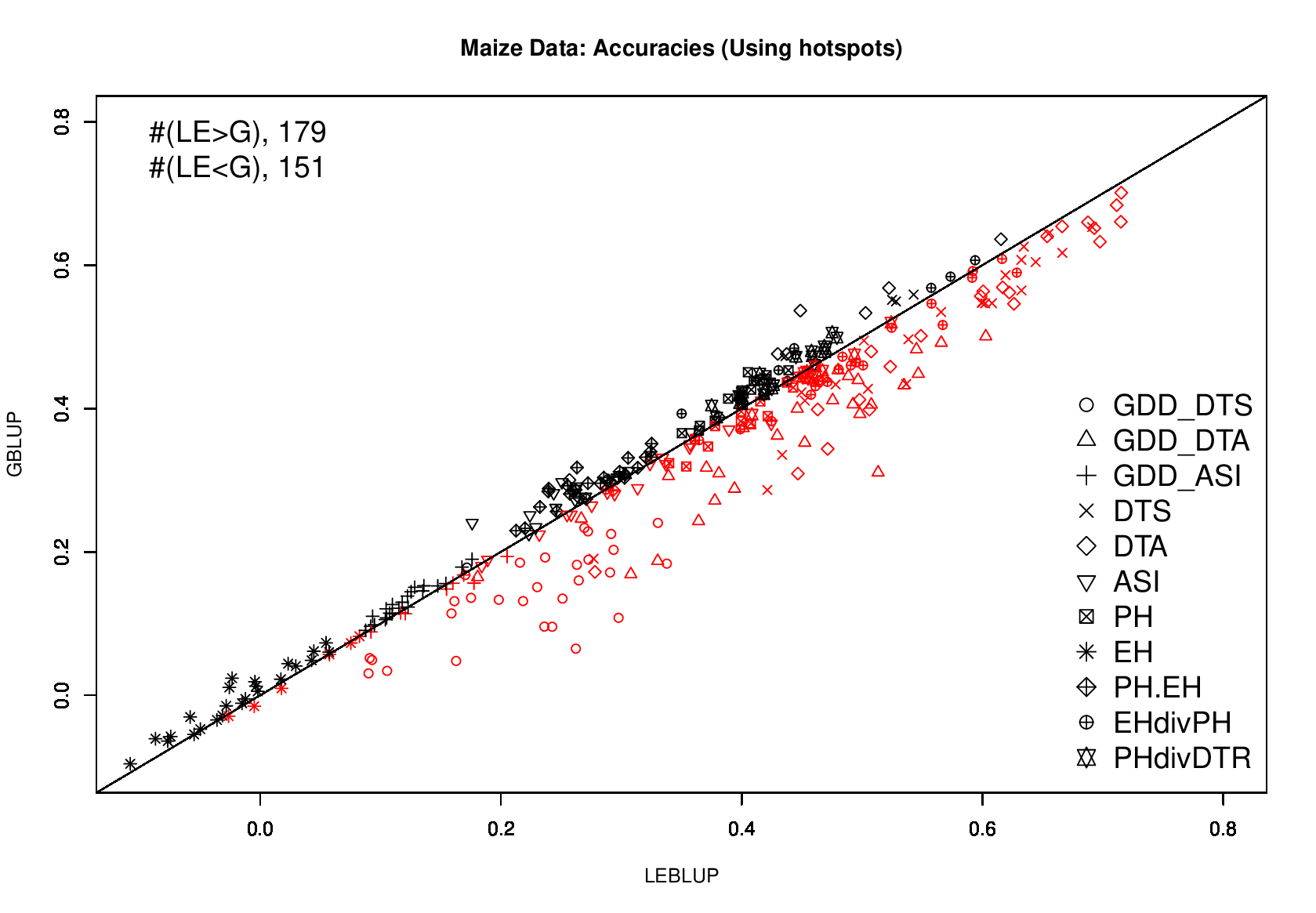} &
    \includegraphics[width=.4\textwidth, angle=270]{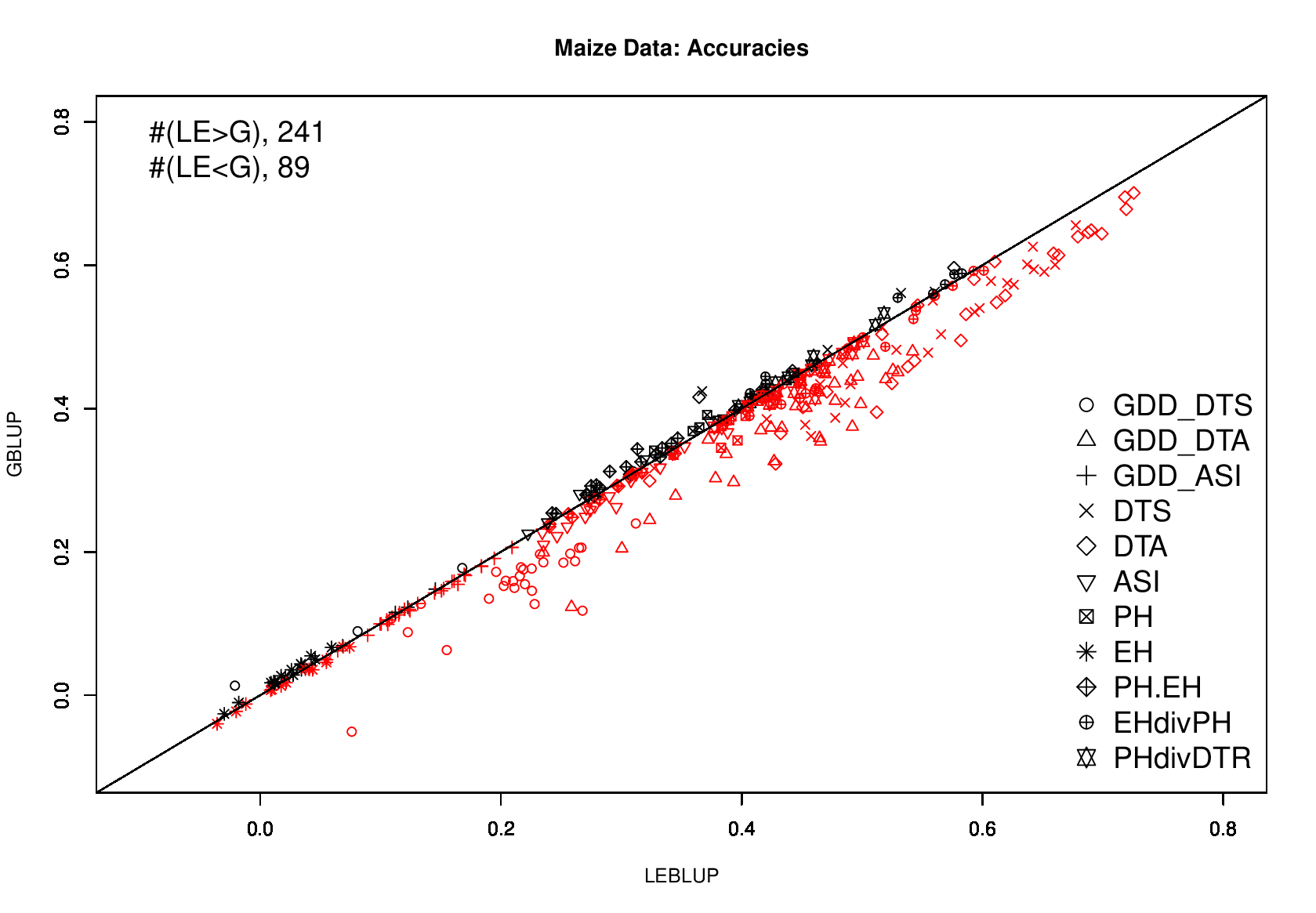}   \\
    \includegraphics[width=.4\textwidth, angle=270]{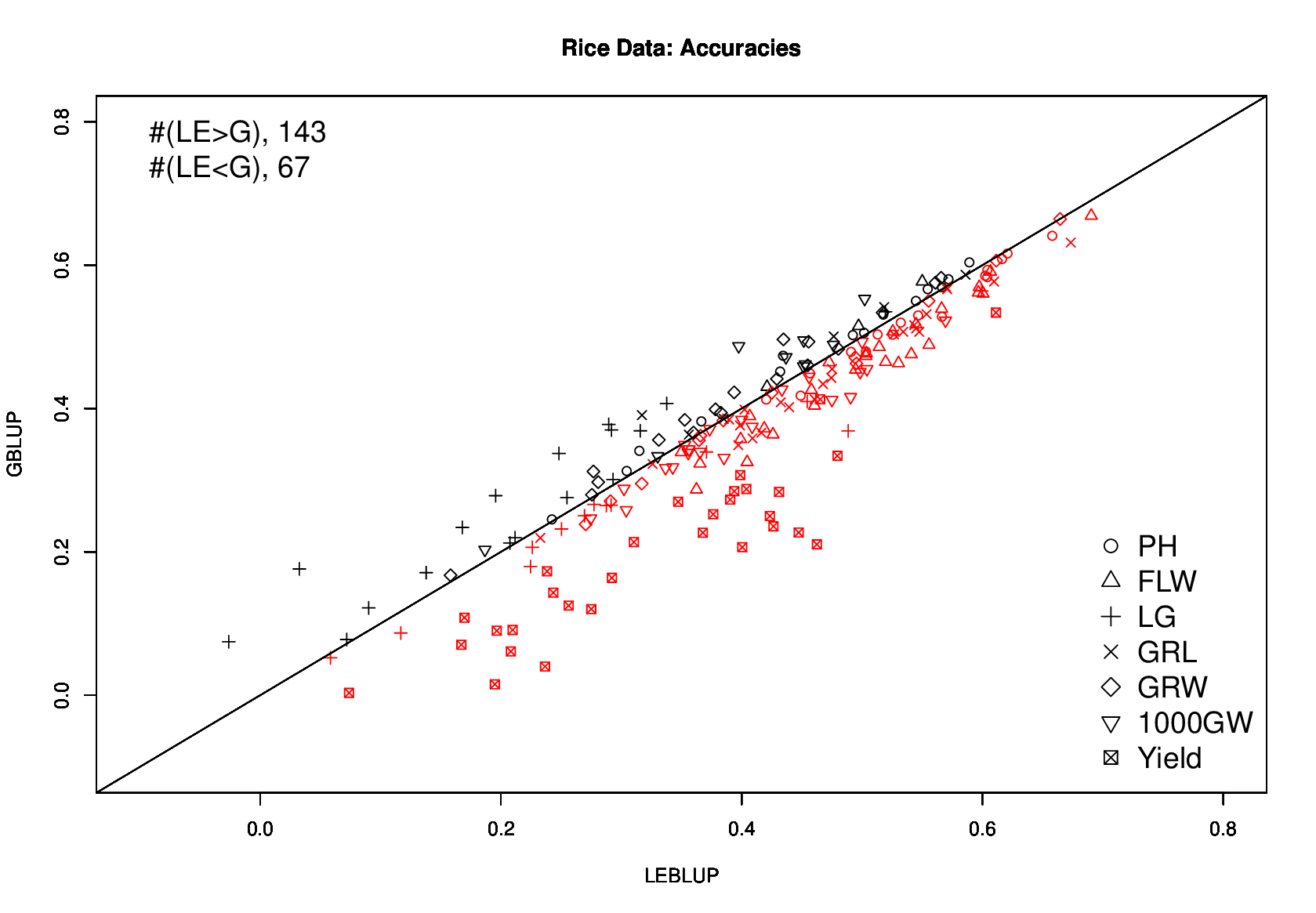} &
    \includegraphics[width=.4\textwidth, angle=270]{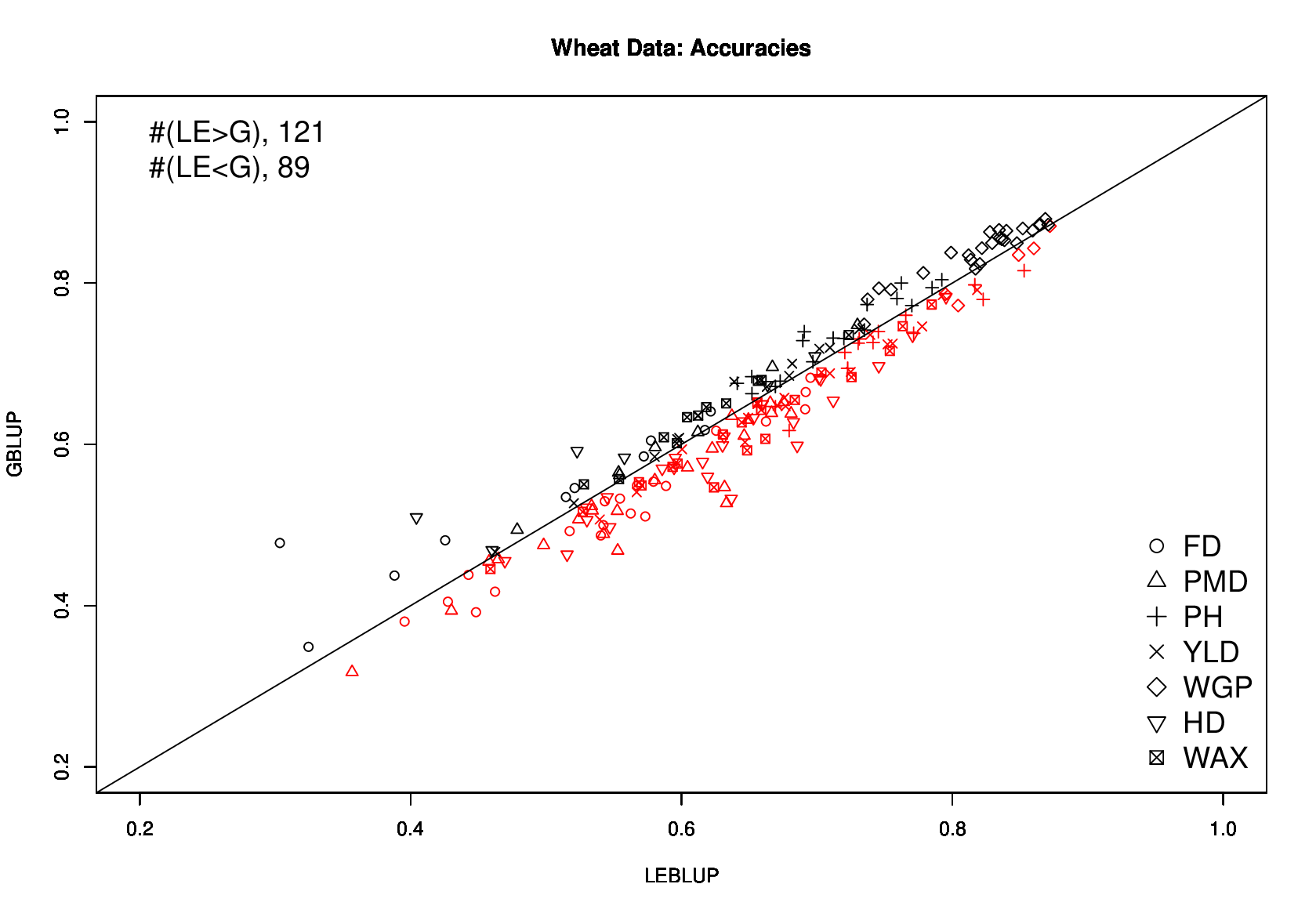}   \\
    \includegraphics[width=.4\textwidth, angle=270]{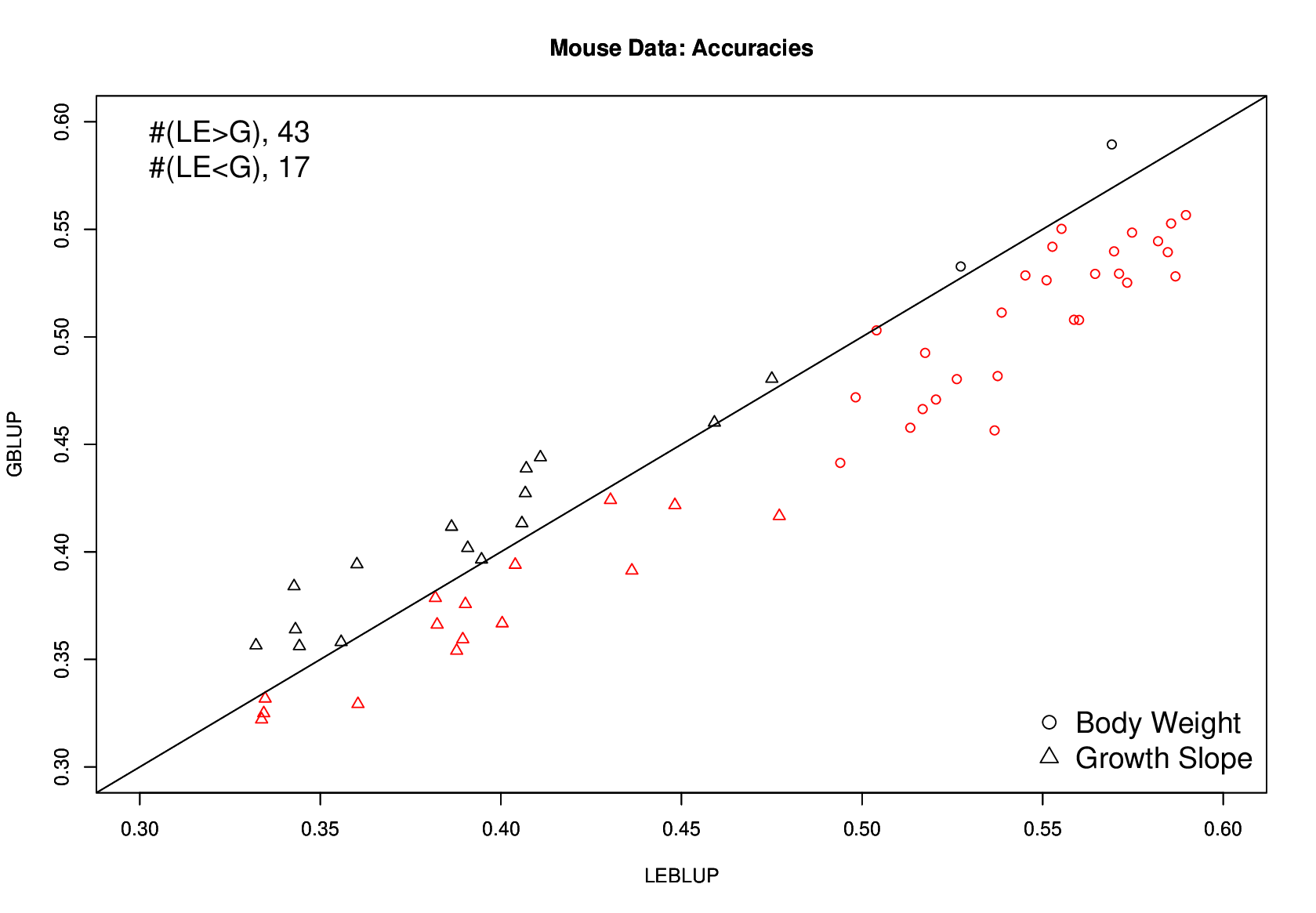} 
  \end{tabular}
  \caption{The accuracies (measured in terms of the correlation between the estimated genetic values and the response variable) for G-BLUP and the LER models compared for Maize, Rice, Wheat and Mouse data sets. The red points below the $y=x$ line are the cases in favor of the LER model.}
    \label{fig:fig5}
\end{figure}

\begin{table}[htb]
\centering
\caption{Summary of the features of the data sets and the hyper-parameter settings for the results presented in Figure \protect\ref{fig:fig5}. Locally epistatic rule approach is much superior.}
\tiny
\label{tab:tab1}
\begin{tabular}{| l | l | l | l | l | l | l | l | l | l | l | l }
\hline
	Data set & \# of Individuals & \# of SNPs & Traits & mean depth & nrules & nsplits & proprow & propcol  \\ \hline
	Rice & 299 & 73K  & PH, FLW, LG, GRL, GRW, & 4 & 500 & 5  & .3 & .1 \\ & & & 1000GW, YLD & & &  &  &  \\ \hline
	Mouse & 1940 & 12K  & Body Weight,  Growth Slope &  & 2000 & 10 & .1 & .05 \\ \hline
	maize  & 4676 & 125K & GDD\_DTS, & 2 & 1000 & 10 & .1 & .05 \\ & & &  GDD\_DTA, GDD\_ASI, DTS, & 2 & 200 & 40 (using hotspots) &  .1 & .05 \\ & & &  DTA, ASI, PH, EH, & &  &  &  & \\ & & & PH.EH, EHdivPH, PHdivDTR &  &  &  &  &  \\ \hline
	wheat & 337 & 3355 & FD, PMD, PH, YLD, WGP, & 1 & 500 & 2 & .3 & .1 \\ & & & HD, WAX &  &  &  & & \\ \hline
\end{tabular}
\normalsize
\end{table}
%------------------------------------------------

For the maize data set, we have used two settings for splitting the markers into contagious and non intersecting regions. In one setting, each chromosome was split into 10 pieces by dividing the chromosome into blocks of approximately the same number of markers. In addition, we have used the recombination hot-spots to split each maize chromosome into 40 pieces.  The rules were extracted using the markers in each region along with the first $3$ PCs of the genome-wide markers. The rice, wheat, and mouse data sets were treated similarly. Table \ref{tab:tab1} contains the settings  used for building the LER models that are presented in the main text of the article. Results for a few other settings are provided in the Supplementary File 1.

%------------------------------------------------

\begin{figure}
\includegraphics[width=10cm, angle=270]{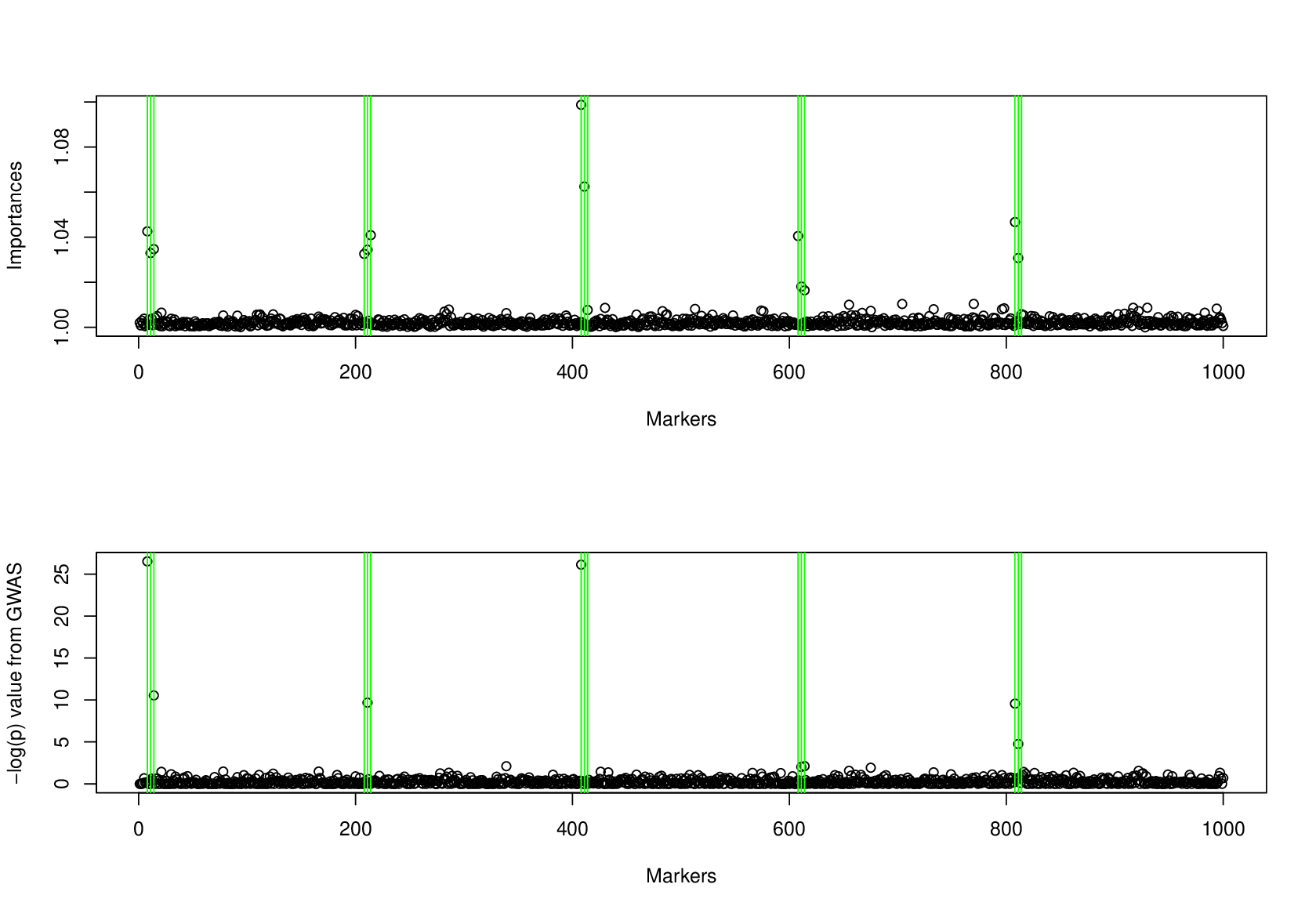}
\caption{We have ran the standard GWAS based on the mixed model and also calculated the importance scores using the trait values and the genotypes generated as described in Table \ref{tab:effects}. The green lines point to the markers that were used to calculate the genetic values. The importance scores and the results from the standard GWAS were similar. More markers were identified correctly as important by the LER approach.}
\label{fig:sim1}
\end{figure}

Mixed models are usually used for association studies. The methods which check interactions usually look for interactions between the markers with the most significant additive effects. This approach can be problematic since the no additive effects might be involved at interacting loci.  In order to show that the LER models can be used to locate interacting loci and to compare it to the standard mixed modeling approach,  we have simulated 1000 independent 0,1,2 coded SNPs for 2000 individuals. We have obtained genetic values for these individuals by generating 5 genetic effects at 5 loci, standardizing them to have variance one and summing them.  Some of these effects were completely additive, some contained marker by marker interactions, marker background interactions or both. The formula for each each of these effects are given in Table \ref{tab:effects}. Half of the individuals were assumed to be males and other half females, which in turn was reflected to the genetic values as a fixed difference of $5$ units.  The final phenotypes for the individulals were obtained by adding iid, zero centered normal random variables to genetic values, the heritability was set to $2/3.$ In Figure \ref{fig:sim1}, a comparison of the association results from a standard additive GWAS approach based on EMMA methodology with the marker importance scores obtained from the LER model. LER model can identify QTL that are missed by ordinary GWAS. 

\begin{figure}
\includegraphics[width=10cm, angle=270]{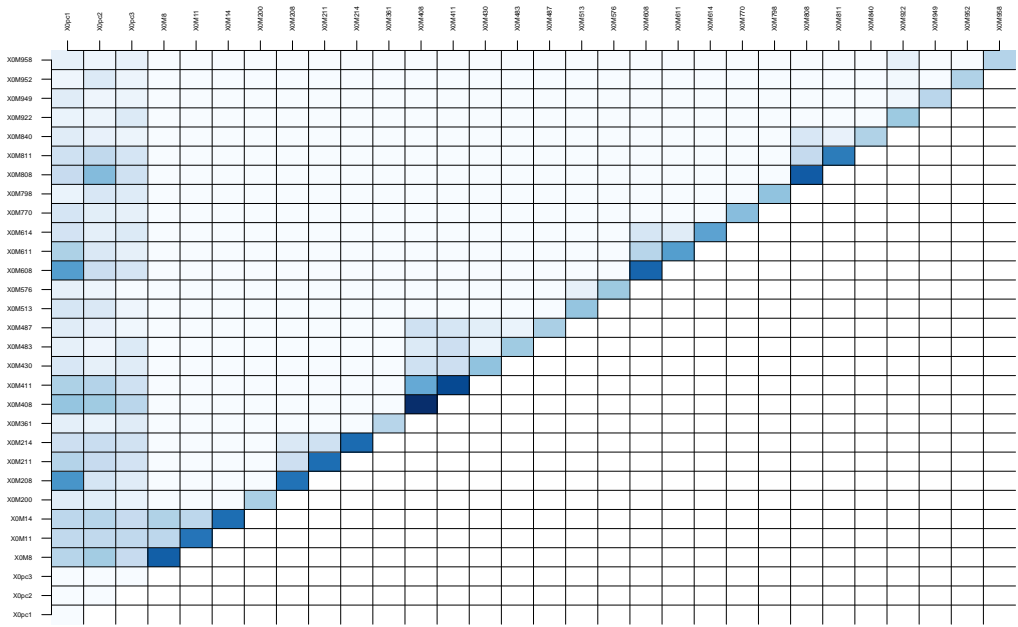}
\caption{The additive and two way interaction importance measures for the most important 30 markers (including the first three principal components) for an instance of data simulated as described in Table \ref{tab:effects}. The darker points point to more important markers, or interactions. Locally epistatic model can find important markers and their interactions.}
\label{fig:sim2}
\end{figure}

\begin{table}[htbp]
  \centering
  \caption{We have simulated 1000 independent 0,1,2 coded snps for 2000 individuals. We have obtained genetic values for these individuals by generating 5 genetic effects at 5 loci (each involving 3 closely located snps), standardizing them to have variance one and summing them.  Some of these effects were completely additive, some contained marker by marker interactions, marker background interactions or both.  Half of the individuals were assumed to be males and other half females, which in turn was reflected to the genetic values as a fixed difference of $5$ units.  The final phenotypes for the individulals were obtained by adding iid, zero centered normal random variables to genetic values, the heritability was set to $2/3.$}

    \begin{tabular}{l}
    \toprule
    \textbf{Effect} \\
    \midrule
    $g_1=(.6*x_{8}+.5*x_{11}-.4*x_{14})$ \\
  $  if (pc_1<0) \left[g_2=.6*x_{208}-.5*x_{211}-.4*x_{214}\right] $  \\ else $\left[g_2=-(.6*x_{208}+.5*x_{211}+.4*x_{214})\right] $\\
  $  g_3=(.6*x_{408}+.5*x_{411}-.4*x_{414})^2$ \\
   $ if(pc_1<0) \left[g_4=((.6*x_{608}+.5*x_{611}-.4*x_{614})^2)\right] $ \\ else $\left[g_4=(-(.6*x_{608}-.5*x_{611}+.4*x_{614})^2)\right] $\\
  $  if(pc_1<0) \left[g_5=((.6*x_{808}+.5*x_{811}-.4*x_{814}+.5*pc_2)^2)\right]$ \\ else $\left[g_5=((-.6*x_{808}-.5*x_{811}+.4*x_{814}-.5*pc_2)^2)\right] $\\
    \bottomrule
    \end{tabular}%
  \label{tab:effects}%

\end{table}%
%----------------------
Figure \ref{fig:sim2} displays the importance and interaction statistics for the most important 20 markers and the first three principal components of the marker data. In addition, for 100 independent replications of the same scenario, we have counted the number of times each of the 15 markers that generate genetic value appear in the top 20 markers selected by each of these methods, these results are summarized in Table \ref{tab:power} and they suggest that LER is superior in identifying QTL.

\begin{table}[htbp]
  \centering
  \caption{Number of times the true loci are recovered by standard GWAS and LER over 100 repetitions of the simulated association experiment described in Table \ref{tab:effects}. }
  \tiny
\begin{tabular}{cccccccccccccccc}
\toprule
Model/Marker & $x_{8}$ & $x_{11}$ & $x_{14}$ & $x_{208}$ & $x_{211}$ & $x_{214}$ & $x_{408}$ & $x_{411}$ & $x_{414}$ & $x_{608}$ & $x_{611}$ & $x_{614}$ & $x_{808}$ & $x_{811}$ & $x_{814}$ \\
\midrule
GWAS &75     & 77     & 68     & 4     & 73     & 74    & 71     & 71     & 1     & 16     & 76     & 25     & 72     & 63    & 0 \\
LER & 96     & 99     & 100     & 100     & 100     & 100     & 100     & 100     & 26     & 100     & 98     & 64     & 100     & 99     & 11 \\
\bottomrule
\end{tabular}
\normalsize
\label{tab:power}
\end{table}

\section{Conclusions}

The focus of this article was building locally epistatic models using rules. However, locally epistatic model building is a general methodology that has three stages.
\begin{enumerate}
\item Divide the genome into regions,
\item Extract locally epistatic effects: Use the training data to obtain a model to estimate the locally epistatic effects. 
\item Post-processing: Combine the locally epistatic effects using an additive model.% fitted in the training data set.
\end{enumerate}

At each of this model building process the researcher would need to make a number of decisions. For example, in all of our implementations of the locally epistatic models we have used non-overlapping contagious regions in this paper. Nevertheless, the regions used in locally epistatic models can be overlapping or  hierarchical. If some markers are associated with each other in terms of linkage or function, it might be useful to combine them together. It is possible to build LER models where each marker defines a region by its neighborhood, this would give overlapping regions. 

The whole genome can be divided physically into chromosomes, chromosome arms or linkage groups. Further divisions could be based on recombination hot-spots or just merely based on local proximity.  We can also use a grouping of markers based on their effects on low-level traits like lipids, metabolites, gene expressions, or based on their allele frequencies. With the development of next-generation sequencing and genotyping approaches, large haplotype datasets are becoming available in many species.  These haplotype frameworks provide substantial statistical power in association studies of common genetic variation across each region.  The locally epistatic framework can be used to take advantage of various annotations of the markers using them to define regions.

The locally epistatic modeling approach overcomes the memory problems that we might incur when the number of markers is very large by loading only subsets of data in the memory at a time. When studying the interactions, an order of magnitude of reduction of complexity can be obtained by only studying the interaction among the blocks instead of interactions among single loci. 

In this paper, we have analyzed 4 real life data sets, and also provided results from simulation studies. In general, our model was very competitive against the G-BLUP model. For some traits the accuracy gains were consistent and considerable, these included for example the yield for the rice data set, body weight for the mouse data set, days to seedling, days to antithesis for the maize data sets, etc,... For many other sets the differences were less significant.  We also presented accuracies for some other settings of the hyper-parameters of the LER algorithm for these data sets in the supplementary file 1. For certain settings, presented there LER model was generally less accurate, however, the improvement on the accuracies were usually robust to small changes in the hyper-parameters. Best settings of the model which will express itself with the best generalization performance that can be estimated via cross-validation or other model selection criteria. These settings in turn might be indicative of the trait architecture. For example, increasing the "mean depth" parameter in wheat data to allow higher order interactions deteriorates the model performance and this can be taken as an indication that for this data set most genetic effects are additive. On the other hand, for the rice data set the best settings for the model have relatively high "mean depths", possibly indicating that in addition to additive effects there is high levels of gene by gene and gene by background interactions in this data set.

The results of the simulated association experiment show that the importance and the interaction scores can be used to identify interesting loci. The comparisons with the standard mixed model based approach showed that LER methodology was superior, it detected loci that weren't detected by the mixed model and at the same time provided a measure of interaction between different types of input variables. We were able to recover most gene by gene and gene by background interactions with the LER model.  We have also described how this methodology can be used to study other forms of interactions. In our belief, the variance components models that use genetic relationship matrices obtained from additive dominance marker codings and their products

Finally we highlight some other strengths specific to the LER models:
\begin{itemize}
\item A method to incorporate marker annotations. 
\item Importance scores for regions, markers, rules are available as a model output.
\item No need to impute the marker data. Model is robust to missing observations in the marker data.  
\item Marker by marker interactions and even higher order interactions are captured and interaction statistics are also available.
\item The model can be used to capture background-gene, or environment-gene interactions.
\end{itemize}

\section*{Acknowledgments}
This research was supported by the USDA-NIFA-AFRI Triticeae Coordinated Agricultural Project, award number 2011-68002-30029.

\bibliographystyle{plain}

\bibliography{LocalEpistasisCitation}

\end{document}